# The Communal Loom: Integrating Tangible Interaction and Participatory Data Collection for Assessing Well-Being

Niti Parikh*
Creative Lead, MakerLAB,
Cornell Tech
ntp27@cornell.edu

Yiran Zhao*
Information Science,
Cornell Tech
yiran.zhao@info.cornell.edu

Maria Alinea-Bravo
ATR-BC, LCAT
NYC H+H/Coler,
maria.alineabravo@nychhc.org

Tapan Parikh
Associate Professor,
Cornell Tech
tsp53@cornell.edu

**ABSTRACT**

For most health or well-being interventions, the process of evaluation is distinct from the activity itself, both in terms of who is involved, and how the actual data is collected and analyzed. Tangible interaction affords the opportunity to combine direct and embodied collaboration with a holistic approach to data collection and evaluation. We demonstrate this potential by describing our experiences designing and using the Communal Loom, an artifact for art therapy that translates quantitative data to collectively woven artifacts.

**CONTEXT**

New York Health + Hospital at Coler (Coler) is a public, long-term rehabilitative care hospital serving mostly low-income and minority patients. It is located on the north end of Roosevelt Island, about two miles away from the Cornell Tech campus on the south end. Craft@Large is an intergenerational, community-based makerspace at Cornell Tech. Residents of Coler Hospital are regular participants in Craft@Large projects and activities, alongside students, senior citizens, design and health professionals and community members.

The idea for the Communal Loom emerged out of a number of Craft@Large sessions involving the co-first authors, who are respectively the the Creative Lead of the Maker Lab, and a Ph.D. student at Cornell Tech, and the third author, a licensed creative arts therapist at Coler, who had extensive experience with designing and instructing art therapy with limited mobility and otherwise differently abled residents at Coler.

The Communal Loom aims to address two key well-being challenges in post-pandemic Coler. The first challenge is to rebuild social connection within and outside Coler. During the pandemic, residents were isolated in their rooms or units for weeks without interacting with external friends/family or other residents at Coler. Such isolation also reduced recreational activities available for residents. The second challenge is for the administration and health care staff to hear and understand the voices of patients. Coler residents had faced a long sequence of challenges, including extended lockdowns, fears about COVID exposure and even the sickness and death of friends and loved ones. This created a general sense of distrust and fear among patients, especially towards Coler administration and other public health officials, who they believed were not considering their safety or needs.

Out of this crucible of challenges emerged the idea of using the Communal Loom as a way of (1) promoting the wellbeing of patients through artistic co-creation, along with (2) recording data about patients' well-being and perspectives. In the rest of this pictorial, we describe the motivation of the Communal Loom, how it was designed and instantiated, and our preliminary experiences using it with patients at Coler Hospital.

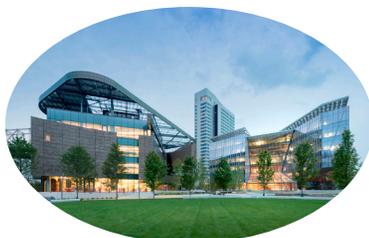
Cornell Tech Campus

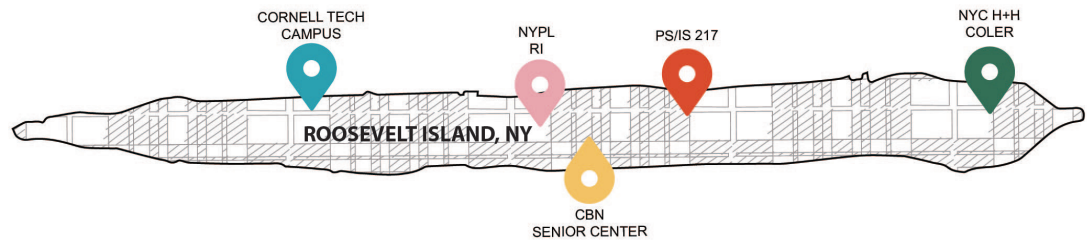
Map of Roosevelt Island with communities partipating in Craft@Large

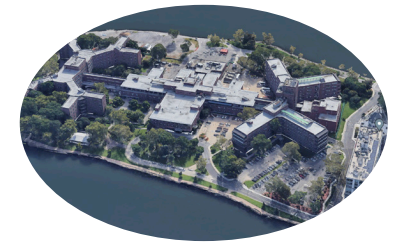
NYC H+H/Coler

## WEAVING AS TANGIBLE SOCIAL INTERACTION AND RECORD KEEPING

Art therapy is a critical part of residential life at Coler. Art therapists host weekly sessions with various forms of artistic creation for individual or group participation. The third author had been using free-form weaving as a regular activity with Coler patients. Weaving fits in the broad category of crafting, which serves "as a purposeful activity that has for decades traditionally been used in occupational therapy and/or rehabilitation" [1]. Weaving as a mode of creativity, social activity has restorative benefits by alleviating sadness in patients, giving them a new sense of purpose and focusing their energy towards a new skill [2].

Weaving on looms or by hands has also provided a platform for record keeping and storytelling for centuries. For example, Inca people used a quipu for monitoring tax obligations, collecting census records, and documenting calendrical information. The cords stored numeric and other values encoded as knots, often in a base ten positional system [3]. Before the 1800s, when the first mechanized Jacquard looms came to existence, weaving has always been a process that combines rote and creative elements, perhaps because, even in its most basic form, the process involves the weaving together of several threads into an integrated fabric. The patterns for complicated weaves were historically recorded on squared paper charts, and later into punch cards for a Jacquard loom, that went on to inform Charles Babbage's Analytical Engine and punch cards used for encoding data in early computers [4].

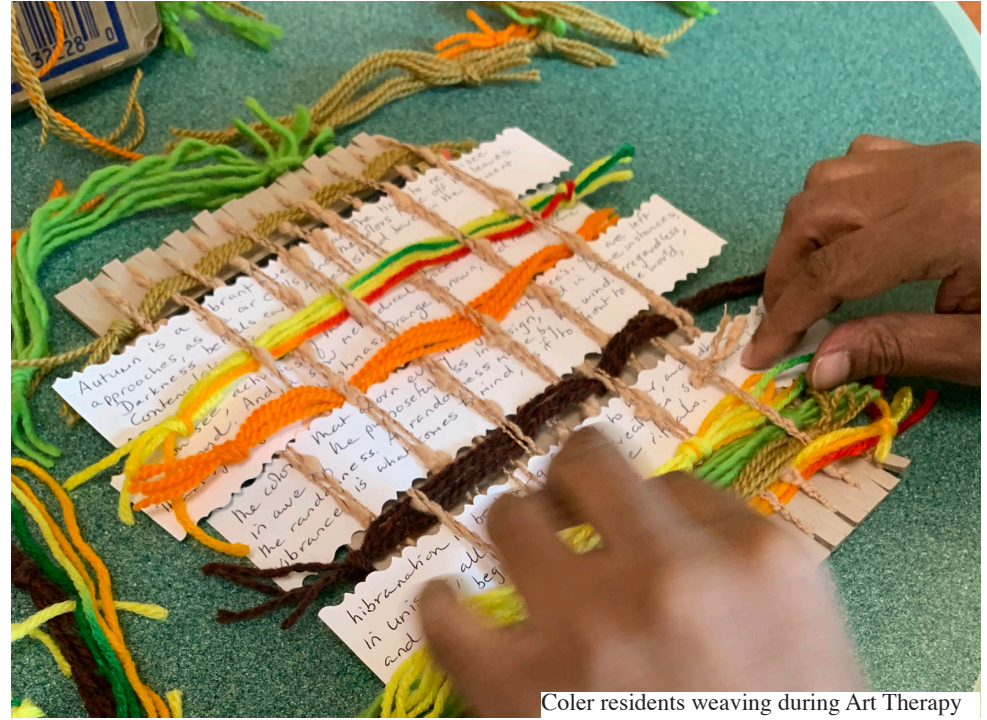
Coler residents weaving during Art Therapy

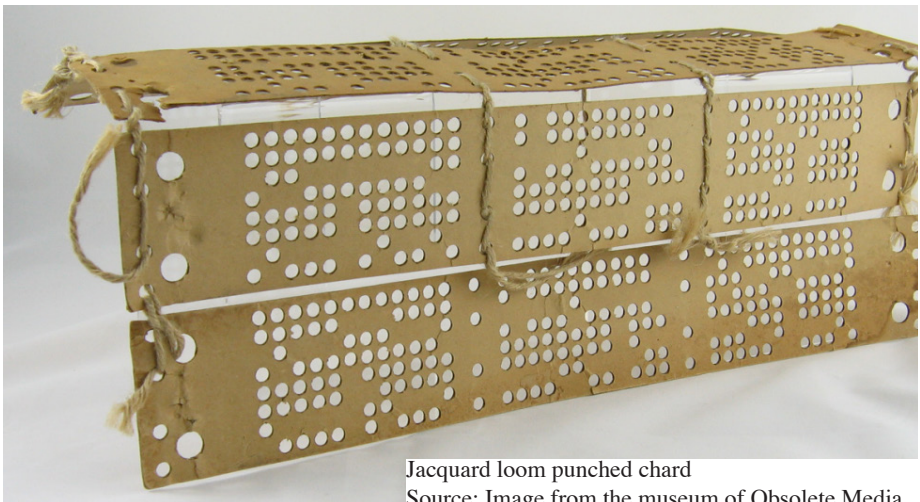
Jacquard loom punched chard
Source: Image from the museum of Obsolete Media

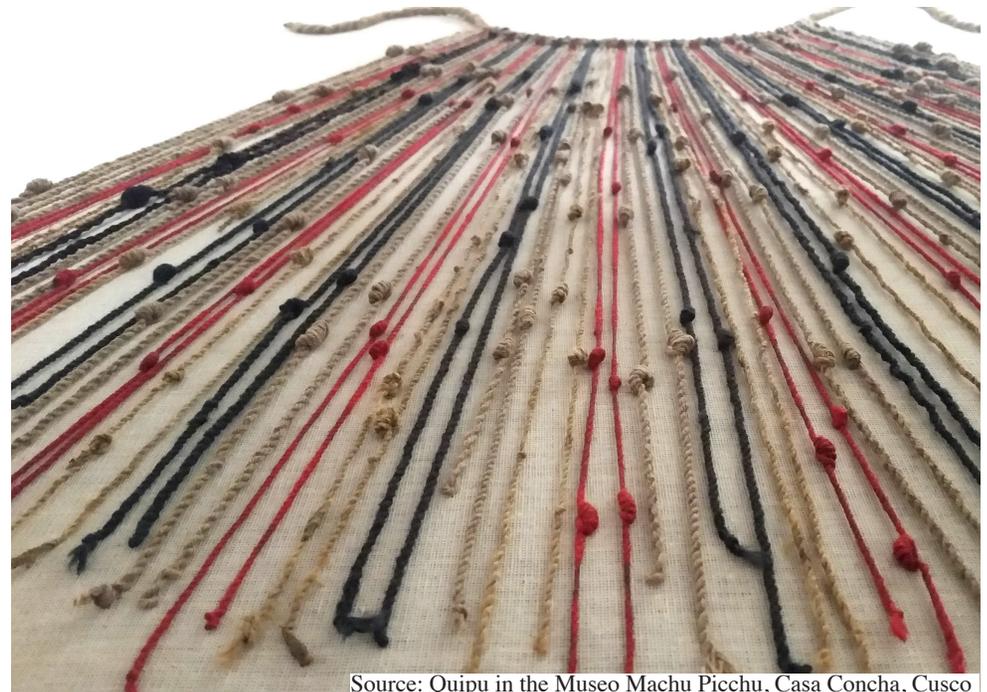
Source: Quipu in the Museo Machu Picchu, Casa Concha, Cusco

## THE COMMUNAL LOOM

The Communal Loom consists of four parts: (1) A questionnaire to gather responses from participants, (2) a bubble sheet-like sheet that translates questionnaire responses into a weaving pattern, (3) rolls of yarn corresponding to possible responses to the questionnaires, and (4) a fixed heddle loom, a loom that is suitable for setting up on a countertop, that we custom made in the lab.

We tested the loom during a regularly scheduled art therapy session at Coler. The art therapist first introduced free form weaving, as previously discussed. Participants chose the yarns spontaneously as their creativity guided them. Then, the art therapist introduced the questionnaires and how questionnaires correspond to the yarn color. The art therapist walked the participants through form-filling and yarn-picking, eventually engaging participants to collectively weave a scroll based on their answers.

Each participant wove eight threads of yarn, corresponding to the answers of each of the survey questions. Between each patient, we included a boundary in a neutral color that allowed us to separate each "row" of data. The final woven piece served as a record of the collective responses of the group on the given day. After the session, we conducted an informal interview with the art therapist about how the process went, and some of the reactions of the patients.

## TENSIONS AND OPPORTUNITIES

Mapping survey questions to weaving patterns allowed us to provide an easy onramp for people who are not used to designing their own weaving patterns. Simply by following the prompts, residents were able to create new patterns based on their responses to each of the queries. In addition, it made it easier for residents to share their feelings and to start new conversations around them.

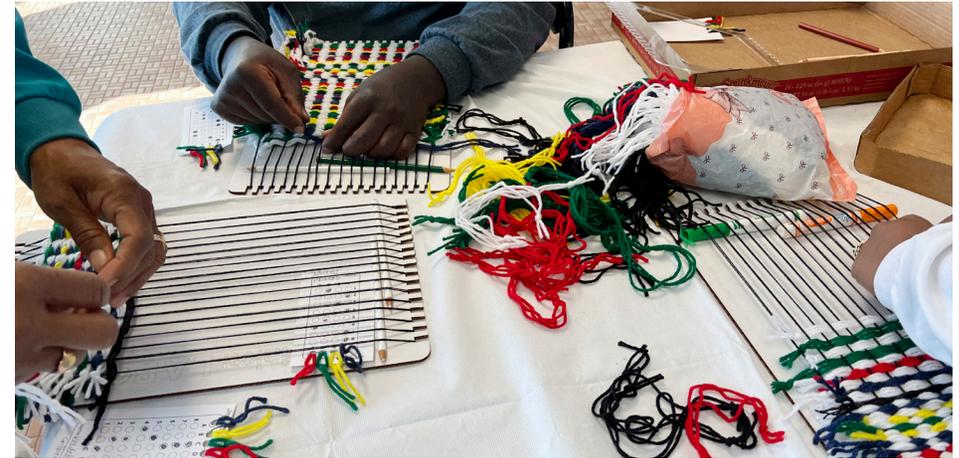
Coler residents weaving with questionnaire

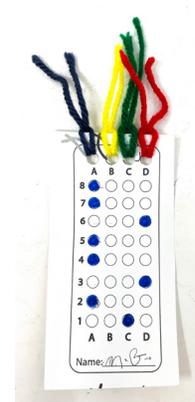
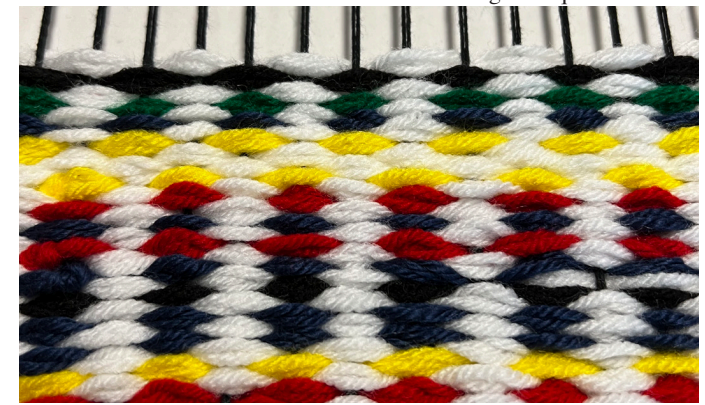

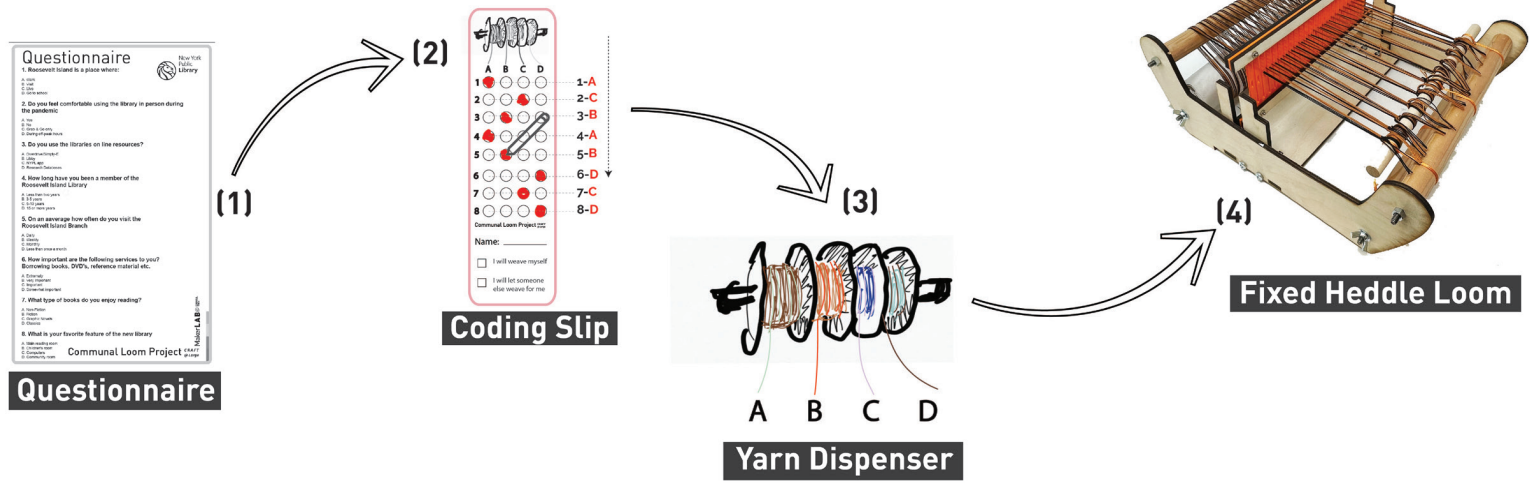
Questionnaire → Coding Slip → Yarn Dispenser → Fixed Heddle Loom

More generally, weaving enabled the integration between artistic creation and data collection. Although designing and mapping a weaving pattern requires expertise, the coded script allowed this process to be made intuitive through a simple mapping, connecting survey responses to color of yarn. The resulting artifact served two functions: as an exhibition of the participants' art and creation, and also as a machine-readable form of data record.

However, some of the participants found that the process became overly mechanical and repetitive, and did not allow them to exercise their own creative independence. This observation allowed us to notice an underlying tension and synergy between repetitive and dynamic processes. Weaving itself is straightforward and repetitive, but the creation of a woven piece is dynamic: participants can choose yarn types and patterns, encompass thoughts and feelings, and thereby project meaning to the woven piece. One of our participants even added a poem to accompany her weaving pattern.

This tension between repetition and creative autonomy is crucial to so many of our social well-being and learning processes. While repetition creates a structure that makes it easy for new participants with limited experience to join easily, we must also create opportunities for individual expression and agency that are inherently unique and individually specific. In the future, we are looking forward to creating new designs, tools and processes that leverage this tension to creatively engage diverse populations in collaborative making activities that serve as moments of collective reflection and individual flourishing.

## ACKNOWLEDGEMENT

We thank the participation of community members and students that made this work possible. We thank the support from NYC H+H/Coler on Roosevelt Island and CBN Senior Center.

## REFERENCES

[1] Vaartio-Rajalin, H., Santamäki-Fischer, R., Jokisalo, P., & Fagerström, L. (2021). Art making and expressive art therapy in adult health and nursing care: A scoping review. International journal of nursing sciences, 8(1), 102-119.

[2] Garlock, L. R. (2016). Stories in the cloth: Art therapy and narrative textiles. Art Therapy, 33(2), 58-66.

[3] Sosa, R., Gerrard, V., Esparza, A., Torres, R., & Napper, R. (2018). DATA OBJECTS: DESIGN PRINCIPLES FOR DATA PHYSICALIZATION.

[4] Lee, V. R., & Vincent, H. (2019). An expansively-framed unplugged weaving sequence intended to bear computational fruit of the loom. In Proceedings of FabLearn 2019 (pp. 124-127).

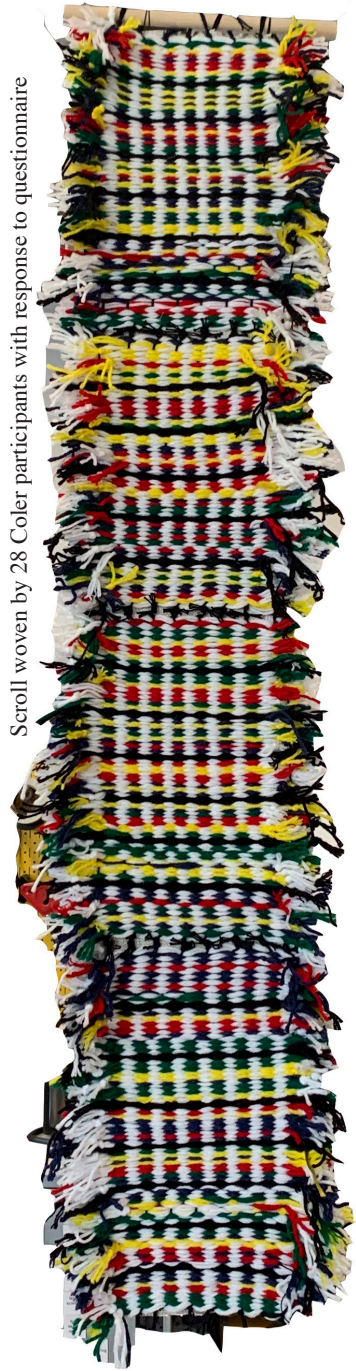

Scroll woven by 28 Coler participants with response to questionnaire

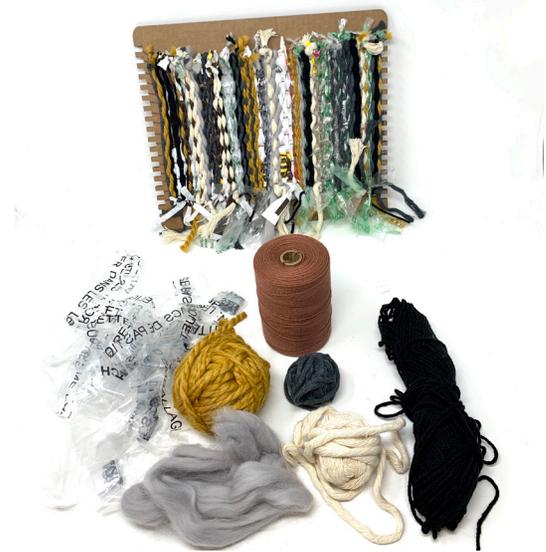

Freeform weaving: Board weaving with a personal selection of weft yarn

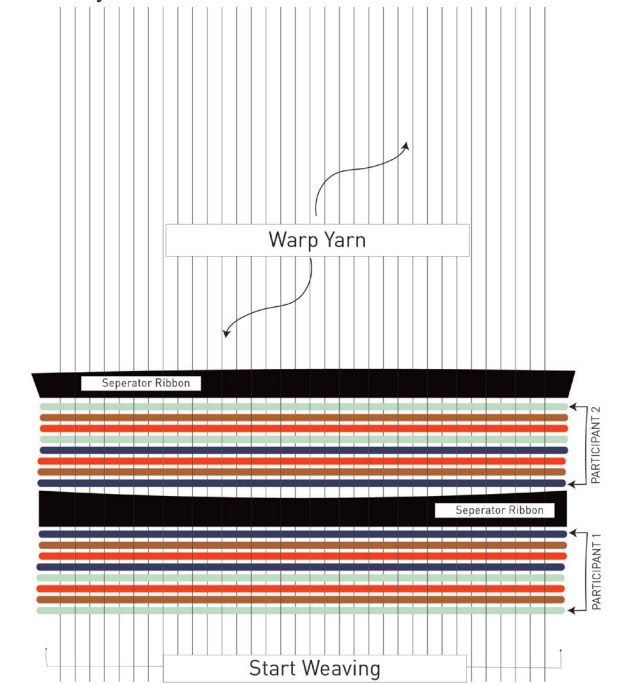

Data weaving: Collective weaving of survey responses by multiple participants